\documentclass[aps,prd,showpacs,twocolumn]{revtex4}
\usepackage{graphicx}
\usepackage{dcolumn}
\usepackage{bm}

\begin{document}

\title{THE XYZs OF CHARMONIUM AT BES}

\author{T.Barnes}

\affiliation{
Physics Division, Oak Ridge National Laboratory,
Oak Ridge, TN 37831, USA\\
Department of Physics and Astronomy, University of Tennessee,
Knoxville, TN 37996,
USA.}

\begin{abstract}
This contribution reviews some recent developments
in charmonium spectroscopy, and discusses related theoretical 
predictions. The spectrum of states, strong decays of states above 
open charm threshold, electromagnetic transitions, and issues 
related to the recent discoveries of the ``XYZ" states are discussed. 
Contributions that BES can make to our understanding of charmonium and related
states are stressed in particular. 

\keywords{charmonium; meson spectroscopy; strong decays; electromagnetic couplings.}
\end{abstract}

\pacs{13.25.Gv, 14.40.Gx, 14.40.Lb}

\maketitle

\section{Introduction}	

The spectroscopy of mesons containing charmed quarks has been the subject 
of intense interest since the discovery of the
surprisingly light and narrow $c\bar s$ candidates 
D$_{s0}(2317)$ 
and
D$_{s1}(2460)$
at BABAR \cite{Aubert:2003fg}
and CLEO \cite{Besson:2003cp} respectively.
These states and other recent developments in charm meson spectroscopy 
have been reviewed by Swanson \cite{Swanson:2006st}.

These discoveries demonstrated that the ``naive quark potential model", 
which assumed that these mesons could be reasonably well described
as $q\bar q$ bound states moving in a simple confining potential
(typically a Coulomb plus linear form, augmented by spin-dependent forces
from one gluon exchange and scalar confinement), was 
much less accurate in some heavy quark systems than had previously been 
thought. For example, the $c\bar s$ scalar D$_{s0}$ state had been 
expected at 2.48 GeV in the potential model of Godfrey and 
Isgur \cite{Godfrey:1985xj}, 
about 160~MeV above the mass of the D$_{s0}(2317)$; previous to 
this discovery, mass discrepancies in the charmed sector were anticipated 
to be perhaps a few 10s of MeVs. It is now widely accepted that the dynamical 
reason for this discrepancy is the strong coupling to the nearby S-wave 
decay channel DK at 2360 MeV, which suggests that the valence ($q\bar q$) 
approximation for hadrons can be misleading, and continuum components 
in charmed meson states may be surprisingly large. 
In the extreme case
this might even suggest DK molecular states \cite{Barnes:2003dj}; an accurate
measurement of the partial width of the radiative transition 
D$_{s1}(2460) \to {\rm D}_s\gamma$, which has been observed by BABAR
\cite{Aubert:2004pw}, may be used to test the conventional $c\bar s$ model.
In all these cases of strongly coupled valence states and hadron continua,
both types of basis states are of course important, and both should be
included in models of the hadronic state vector.

These discoveries dramatically illustrate the crucial importance 
and richness of experiments on the spectroscopy of heavy-flavor hadrons, 
where striking and unexpected discoveries have been made repeatedly in 
recent years.

\section{Charmonium}

Charmonium spectroscopy has also been a very active topic recently.
The realization that studies of B decays could make important
contributions to charmonium spectroscopy \cite{Eichten:2002qv}
was followed by the discovery 
of the remarkable X(3872) state 
by the Belle Collaboration at KEK \cite{Choi:2003ue}, 
in the final state $J/\psi \pi^+\pi^-$.
Although it was initially thought that this 
might be one of the as yet undiscovered narrow D-wave $2^-$ 
states, the mass and width were found to be inconsistent
with this assignment~\cite{Barnes:2003vb}. 

The near degeneracy
of the X with the mass of a neutral D$^0$D$^{*0}$ pair 
suggested that this might instead be an S-wave DD$^*$ molecule, 
strongly isospin violating since it would be largely a neutral 
pair \cite{Swanson:2003tb,Close:2003sg}.
One would expect a weakly bound DD$^*$ system if bound by pion 
exchange \cite{Tornqvist:2003na}
to have J$^{PC} = 1^{++}$ quantum numbers, and there is now much
evidence that this J$^{PC}$ assignment is correct \cite{Abe:2005iy}.
The recent observation of the X(3872) with comparable
strength in the two modes $J/\psi \rho^0$ {\it and} $J/\psi \omega$ 
\cite{Abe:2005ix} is the most striking evidence of the validity of this 
D$^0$D$^{*0}$ charm molecule model. 
Thus the early speculations \cite{Voloshin:1976ap,DeRujula:1976qd}
that there might be charmed meson molecules appear to be confirmed, 
although not the $\psi(4040)$ as was originally suggested.

\begin{figure}[ht]
\vskip 0.9cm
\includegraphics[width=9.0cm,angle=0]{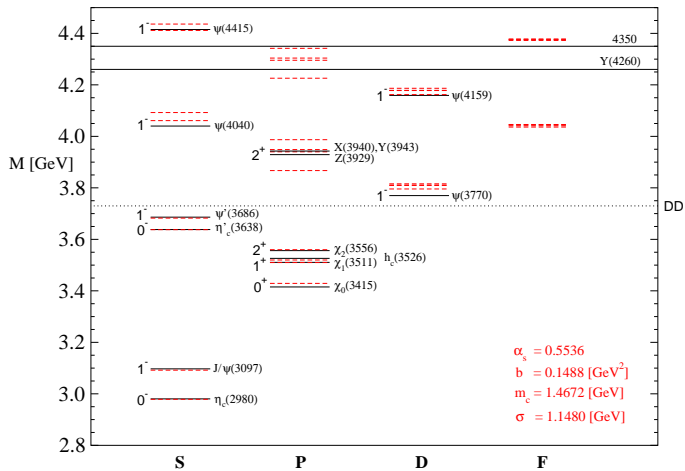}
\caption{
The current experimental status of charmonium (and possible charmonium
hybrid) spectroscopy, compared to the predictions of a nonrelativistic potential model.
Experimental levels are solid lines, and theoretical levels are dashed. The open-charm
threshold at 3.73~GeV is also shown.}
\vskip -0.5cm
\label{fig1}
\end{figure}

The surprises in charm-strange meson spectroscopy and the discovery of the 
X(3872) motivated several recent detailed theoretical studies of charmonium 
spectroscopy \cite{Barnes:2005pb,Eichten:2005ga}. 
The known spectrum of charmonium 
candidates has until recently been in remarkably good agreement with 
potential model predictions;  Fig.~\ref{fig1} for example shows the charmonium 
spectrum in a Coulomb plus linear potential model (abstracted from
Ref.\cite{Barnes:2005pb} and updated) compared to experiment.  
Studies of the strong decays of states above the open-charm 
threshold of 3.73 GeV \cite{Barnes:2005pb,Eichten:2005ga} showed that
in addition to the narrow $2^{-+}$ and $2^{--}$ states, the 
$3^{--}$ $^3$D$_3$ $c\bar c$ state should also be quite narrow ($< 1$ MeV)
due to the large F-wave centrifugal barrier against its decay mode DD.
All of the allowed open-charm strong decay amplitudes and E1 electromagnetic 
partial widths of the $c\bar c$ states up to 4.42~GeV were evaluated 
in these theoretical studies, and future comparisons of these predictions 
with experimental results will provide interesting tests of our understanding 
of the physics of charmonium.

We note in passing that there are more fundamental studies of charmonium
using lattice QCD, which thus far have found results for the spectrum of
states that are quite similar to the predictions of potential models.
This work was reviewed at this meeting by Morningstar \cite{Morningstar}. 

\subsection{The XYZ States Near $3.9$ GeV }

Three of the states discovered in recent experimental studies, 
the X(3943), Y(3940) and Z(3930), have masses roughly consistent with 
expectations for 2P states (radial excitations of the $\{ \chi_J\} $)
and perhaps the 3S $\eta_c''$; see Fig.~\ref{fig1}. 
These obvious $c\bar c$ assignments should be explored in detail before more 
exotic interpretations such as molecules or anomalously light 
$c\bar c$ hybrids are seriously entertained. 

\begin{table}[hb]
\caption{Allowed open-charm decay modes and partial widths ($^3$P$_0$ model)
of $C=(+)$ 2P $c\bar c$ states.}
\vskip 0.5cm
{\begin{tabular}{@{}cccc@{}} \toprule
State & Quantum Numbers & Mode  & Width   \\
&  &  & (MeV)\\ \colrule
$\chi_2'(3929)$ & 2$^3$P$_2$ ($2^{++}$) & DD$^*$           & 11.3 \\
                &                       & DD\phantom{$^*$} & 34.3 \\
$\chi_1'(3940)$ & 2$^3$P$_1$ ($1^{++}$) & DD$^*$           & 140. \\
$\chi_0'(3940)$ & 2$^3$P$_0$ ($0^{++}$) & DD\phantom{$^*$} & see text \\
\botrule
\end{tabular} 
\label{ta1}
}
\end{table}

\eject
Since the only open-charm strong decay modes available these states 
are DD and DD$^*$, a simple comparison of the states observed in
these two decay modes can provide valuable information. 
The predicted partial widths of 2P states into these modes 
in the $^3$P$_0$ decay model of Ref.\cite{Barnes:2005pb}, 
generalized to the masses indicated, 
are given in Table~\ref{ta1}. Note that the width of the 
$\chi_0'$ is problematic, as there is a node in the $^3$P$_0$ model DD 
decay amplitude near the physical point.  

\subsubsection{Z(3930)}

Of the three new XYZ states, the Z(3930) and its proposed assignment
should be the easiest to test in future experiments.
This state was reported by the 
Belle Collaboration \cite{Uehara:2005qd} in $\gamma\gamma$ collisions, 
in the processes $\gamma\gamma \to {\rm Z}(3930) \to {\rm D}^+{\rm D}^-$
and ${\rm D}^0\bar {\rm D}^0$. Belle suggested that this might be the 
radially excited J=2 $\chi_2'$, since there is a preference for J=2 in the 
DD angular distribution.

The reported strength of the combined $\gamma\gamma$ and DD couplings 
is indeed roughly consistent with this $\chi_2'$ assignment.
The published \cite{Uehara:2005qd} Belle results are
\begin{equation}
M = 3929 \pm 5 \pm 2~MeV,
\end{equation}
\begin{equation}
\Gamma = 29~\pm~10~\pm~2~MeV
\end{equation}
and
\begin{equation}
\Gamma_{\gamma\gamma} \cdot B_{DD}\Big|_{expt.} = 0.18 \pm 0.05 \pm 0.03~keV.
\end{equation}
In comparison, the quark model predicts a two-photon width for a $\chi_2'$
of about $\Gamma_{\gamma\gamma} = 0.64$~keV \cite{Barnestwophot}
(M\"unz \cite{Munz:1996hb} quotes theoretical results for this number
from several models, which give $\Gamma_{\gamma\gamma}  = 0.317-0.684$~keV), 
and a DD branching fraction of about 75\%.
(This DD branching fraction is from the $^3$P$_0$ decay model of 
Ref.\cite{Barnes:2005pb}, generalized to a $\chi_2'$ mass of 3.929 GeV.) 
Combining the $\Gamma_{\gamma\gamma}$ range quoted by M\"unz and the 
predicted DD branching fraction gives the theoretical result
\begin{equation}
\Gamma_{\gamma\gamma} \cdot B_{DD}\Big|_{theor.}  = 0.24-0.51~keV.
\end{equation}
Given the uncertainties in these calculations, this may be regarded 
as rough agreement 
between theory and experiment for a $\chi_2'$. The definitive 
test of this assignment would be the observation of a DD$^*$ mode; 
the expected relative branching fraction is DD$^*$/DD = 0.35, 
and the only plausible competing assignment, $0^{++}$ 2$^3$P$_0$, 
does not lead to a DD$^*$ final state. (The $1^{++}$ 2$^3$P$_1$ 
of course cannot be made in $\gamma\gamma$ collisions.)

\eject
\subsubsection{X(3943)}

The X(3943) was reported by Belle \cite{Abe:2005hd} in the double 
charmonium production reaction $e^+e^- \to J/\psi\; {\rm X}(3943)$
in the final state DD$^*$, in both charged and neutral modes. 
The fitted mass and width are
\begin{equation}
M = 3943 \pm 6 \pm 6~MeV,
\end{equation}
\begin{equation}
\Gamma = 15 \pm 10~MeV~\hbox{or}~<~52~MeV\ (90\% \ c.l.).
\end{equation}
Since the only other charmonium states seen recoiling against the $J/\psi$ 
with comparable strength in this (poorly understood) process are the 
$\eta_c$, $\chi_0$ and $\eta_c'$, the obvious assignment for this state 
is $\eta_c''$. ($\chi_0'$ cannot decay to DD$^*$). 

The reported total width however is surprisingly small for an $\eta_c''$
assignment; one expects $\Gamma = 70~MeV$ in the $^3$P$_0$ decay model, 
using the reported mass as input. Of course the experimental total width 
is not very well determined, and the discrepancy 
may disappear with better statistics. The mass is also surprising 
for an $\eta_c''$, since it is about 100~MeV below the presumably 
3$^3$S$_1$ partner $\psi(4040)$; in the 2S states, the $\psi'-\eta_c'$ 
splitting in contrast is only about 30 MeV. If the X(3943) is indeed the 
$\eta_c''$, either the mass is not yet accurately determined, or there are 
important mass shifts in the 3S states relative to 2S. Testing the 
$\eta_c''$ assignment is a simple matter of establishing whether 
the angular distribution of DD$^*$ final states is P-wave  
(J$^P = 0^-$); alternative J=1 and J=2 2P assignments lead to 
S- and D-wave DD$^*$ final states.

\subsubsection{Y(3940)}

This may be the least well established of the new 
XYZ states. Evidence for this state was reported by
Belle \cite{Abe:2004zs} as an $\omega J/\psi$ threshold 
enhancement in the charged B decays
B$^{\pm}\to {\rm K}^{\pm}\omega J/\psi$.
Assuming that this was
due to a resonance, Belle quoted a mass and width of
\begin{equation}
M = 3943 \pm 11 \pm 13~MeV,
\end{equation}
\begin{equation}
\Gamma = 87 \pm 22~MeV.
\end{equation}

Of course the observation of a charmonium state
in a closed-charm final state such as $\omega J/\psi$
with a relatively large branching fraction
($B_{B\to KY}\cdot B_{Y\to \omega J/\psi} = 7.1\pm 1.3\pm 3.1 \cdot 10^{-5}$)
is very surprising, since the corresponding close-charm decay partial width
for $\psi' \to J/\psi \pi\pi$ is only about 140~keV. Since the Y(3940)
has a total width near 100~MeV, one might expect an $\omega J/\psi$ 
branching fraction of roughly $10^{-3}$. Since the known 
{\it total} B meson 
branching fractions to the 1P $c\bar c$ states such as the $\chi_1$ 
are only an order of magnitude larger,
for example $B_{B\to K^+ \chi_1} = 5.3\pm 0.7 \cdot 10^{-4}$, the reported
Y(3940) signal appears to imply an anomalously large branching fraction
for Y$(3940) \to \omega J/\psi$. Either the Y(3940) is quite unusual 
in populating this decay mode, or it is not actually due to a resonance.

The mass, width and $\omega J/\psi$ decay mode of this state, and the 
fact that the $2^{++}$ 2P state is likely the Z(3930), suggest that 
the least implausible
$c\bar c$ assignment for the Y(3940) is $1^{++}$ 2$^3$P$_1$. This state is
predicted to have a total width of about 140~MeV, dominantly into the open-charm
mode DD$^*$. A search for this signal in DD$^*$, with a much larger 
branching fraction than $\omega J/\psi$, is the obvious test of this 
assignment. If this assignment is correct, the closed-charm mode 
$\omega J/\psi$ may have come about through an inelastic final state
interaction, Y$(3943) \to ({\rm DD}^*, {\rm D}^*{\rm D}^*) \to \omega J/\psi$. 
The fact that these are near-threshold S-wave processes would 
enhance this FSI effect.

\section{Future BES Contributions}

As an $e^+e^-$ machine in the E$_{cm}\approx 3$-$4$~GeV mass range,
the contribution of BES to these studies of charmonium and related states
will usually involve the initial production of a $1^{--}$ resonance. The four 
known $1^{--}$ states with relatively large $e^+e^-$ couplings are the 
$\psi(3770)$, $\psi(4040)$, $\psi(4160)$ and $\psi(4415)$. As the physics 
questions that can be addressed using each of these entry states differ 
somewhat, we will briefly discuss them individually here.

\subsection{$\psi(3770)$} 
 
The $\psi(3770)$ is a well established, dominantly D-wave $c\bar c$ state. 
One topic of interest here is the level of $^3$S$_1$-$^3$D$_1$ mixing
present in the $\psi(3770)$; a moderate mixing is required to explain the
$e^+e^-$ width of this state, since a pure D-wave $c\bar c$ would have 
a much smaller $e^+e^-$ width than is observed. There is a cross-check 
of this mixing angle using E1 radiative widths; the partial width for 
$\psi(3770) \to \gamma \chi_2$ is very sensitive to this mixing 
angle \cite{Barnes:2005pb,Ding:1991vu},
and recent upper limits from CLEO-c are rather close to the expected 
width \cite{Briere:2006ff}.
It would be very interesting for BES to measure this E1 width. The remaining
transitions $\psi(3770) \to \gamma \chi_{0,1}$ are also interesting as checks 
of the theory, but are much less sensitive to this mixing angle.

Another very interesting question that can be addressed using the 
$\psi(3770)$ is the strength of the coupling of orbitally excited 
$c\bar c$ states to $p\bar p$. This is a very important question for the 
future PANDA experiment \cite{PandaTechnicalProgress} at GSI, 
which plans to use $p\bar p$ annihilation to 
produce excited $c\bar c$ and charmonium hybrids. At present we have the
intriguing experimental observation that the L=1 $c\bar c$ $\chi_J$ states
couple much more strongly to $p\bar p$ than the L=0 $J/\psi$, but whether
this trend continues to L=2 is an open question. BES can easily answer this 
question through a high-statistics search for $\psi(3770) \to p\bar p$.
Three body decays such as $\Psi \to p\bar p m $ (where $\Psi$ is a generic 
charmonium or charmonium hybrid resonance and $m$ is a light meson)
are also very interesting in this regard, and can be used to estimate the
associated production cross section for $p \bar p \to m \Psi $ (see
Ref.\cite{Lundborg:2005am}); this reaction will be used by PANDA
to search for J$^{PC}$-exotic charmonium hybrids. 

\subsection{$\psi(4040)$ and $\psi(4160)$} 
 
The $\psi(4040)$ and $\psi(4160)$ have important applications in the 
study of the open-flavor strong decay mechanism, 
and are also of interest because their radiative transitions can be used 
to access lower-mass $c\bar c$ candidates such as the new XYZ states.

\subsubsection{Strong Decay Studies using D$^*$D$^*$}

The strong decays of the vectors $\psi(4040)$ and $\psi(4160)$ 
to D$^*$D$^*$ are especially interesting, since this is their only 
``multiamplitude" decay mode. The decays to DD and DD$^*$ are 
single amplitude decays, respectively $^1$P$_1$ and $^3$P$_1$, 
so one learns nothing new about the decay process by studying 
their angular distributions. The decays to D$^*$D$^*$ however have
three allowed amplitudes, $^1$P$_1$, $^5$P$_1$ and $^5$F$_1$, and an 
experimental determination of the ratios of these amplitudes can be
used as an important test of the decay model, specifically of the 
quantum numbers of the light $q\bar q$ pair produced 
in the decay. The two principal models assumed by theorists to study 
these $c\bar c$ decays at present are the $^3$P$_0$ (scalar) 
model \cite{Barnes:2005pb} and the Cornell (timelike vector) 
model \cite{Eichten:2005ga}; these give different predictions for 
the relative D$^*$D$^*$ decay amplitudes, which have not been tested 
experimentally. Of course these quantum numbers are not especially 
fundamental, and many other possibilities can be imagined. 
In the $^3$P$_0$ model, the P-wave $\psi \to {\rm D}^*{\rm D}^*$ decay amplitudes 
have simple ratios \cite{Barnes:2005pb} that are independent of the radial 
wavefunction,
$^5$P$_1$/$^1$P$_1 = -2\sqrt{5}$ for an initial S-wave $\psi$ state and
$^5$P$_1$/$^1$P$_1 = -1/\sqrt{5}$ for an initial D-wave. An S-wave
$\psi$ gives a vanishing $^5$F$_1$ D$^*$D$^*$ amplitude, whereas for
a D-wave it is large. (For a pure D-wave $\psi(4160)$ this 
$^5$F$_1$ D$^*$D$^*$ amplitude is predicted to be dominant.)

If BES can measure these amplitude ratios, this important information 
will allow theorists to formulate more accurate models of $c\bar c$
strong decays, and should greatly improve our understanding of this dominant 
QCD strong decay process generally. 

\subsubsection{Accessing the XYZ States}

\begin{table}[h]
\caption{Theoretical E1 radiative partial widths of the 
$\psi(4040)$ and $\psi(4160)$ into $C=(+)$ 2P $c\bar c$ states.}
\vskip 0.5cm
{\begin{tabular}{@{}ccccc@{}} \toprule
Initial State & Final State & E1 Width  & E1 B.F.      & \\
              &             & (keV)     &              & \\ \colrule
$\psi(4040)$ & $\chi_2'(3929)$  & $56.$ & $0.7 \cdot 10^{-3}$   &  \\
             & $\chi_1'(3940)$  & $25.$ & $0.3 \cdot 10^{-3}$   &  \\
             & $\chi_0'(3940)$  & $8.3$ & $0.1 \cdot 10^{-3}$   &  \\
&  &  &  & \\
$\psi(4160)$ & $\chi_2'(3929)$  & $9.9$  &  $0.1 \cdot 10^{-3}$  &  \\
             & $\chi_1'(3940)$  & $129.$ &  $1.3 \cdot 10^{-3}$  &  \\
             & $\chi_0'(3940)$  & $172.$ &  $1.7 \cdot 10^{-3}$  &  \\
\botrule
\end{tabular}
\label{ta2}
}
\end{table}

The $\psi(4040)$ and $\psi(4160)$ can be used as $1^{--}$ entry states 
for the study of the new XYZ states near 3.9~GeV. As shown in 
Table~\ref{ta2},  
both these states are expected to have relatively large E1
branching fractions into the 2P $c\bar c$ multiplet, 
$\psi(4040,4160) \to \gamma \chi_J'$. 
(These E1 partial widths were calculated as in Ref.\cite{Barnes:2005pb},
for the masses given in the table.)
This may allow the identification of the 2P resonances 
through their subsequent hadronic decays. In this approach one would study the 
invariant mass and angular distributions of the final charmed mesons in the 
processes $e^+e^- \to \psi(4040,4160) \to \gamma {\rm DD}$ and $\gamma {\rm DD}^*$.  

\subsection{States above 4.2 GeV}

The recently discovered $1^{--}$ states Y$(4260)$ \cite{Aubert:2005rm,Coan:2006rv}
and the (post conference) new state at
$4350$~MeV \cite{Ye}
are strictly speaking not within the assigned topic of this talk; 
since the previously known $1^{--}$ states $\psi(4040)$, $\psi(4160)$ and $\psi(4415)$ 
fill the available $1^{--}$ $c\bar c$ assignments to just above 4.4~GeV, 
these two new states appear unlikely as $c\bar c$ candidates. They have been 
reported only in the closed-charm modes 
$J/\psi \pi\pi$ and $\psi'\pi\pi$ respectively, which 
are naively expected to be very weak. (Of course there is a LGT prediction that 
hybrids might preferentially populate these modes \cite{McNeile:2002az}.) 
Here the most important task is 
probably to search for these states in all accessible 
open-charm modes, which might be expected to be
dominant even in hybrids. 

Finally, the highest-mass $c\bar c$ state currently known is the $1^{--}$ $\psi(4415)$,
which is usually given a 4$^3$S$_1$ assignment. Nothing is currently known about its
exclusive decay modes. (The PDG \cite{PDG2006} says that the $\psi(4415)$ decays 
dominantly to ``hadrons", which is not especially surprising.) Calculations of 
the decay branching fractions of a 4$^3$S$_1$  $c\bar c$ $\psi(4415)$ 
in the $^3$P$_0$ model \cite{Barnes:2005pb} predict that the largest mode should be 
the unusual DD$_1$, and in pure D-wave rather than S-wave! It would clearly be 
a very interesting test of strong decay models to
measure the strong decay amplitudes and branching fractions of this state. There is also
an ``industrial" application of the $\psi(4415)$; by running on the high mass tail of 
this resonance, one can expect a relatively large branching fraction into the enigmatic
D$_{s0}(2317)$ \cite{Barnes:2005pb,Guo:2005cs}, which otherwise is very difficult to produce 
with useful statistics. 
A study of interesting decays such as the radiative branching fraction
of the D$_{s0}(2317)$ into $\gamma $D$_s^*$ could then be carried out at BES;
this would be valuable in determining the relative size of the $c\bar s$ and DK
components of the D$_{s0}(2317)$.

\section*{Acknowledgments}

I would like to thank the organisers of Charm2006 for their kind invitation to
present this material, and for the opportunity to discuss 
charm and charmonium physics with my fellow participants.
This research was supported in part by the 
U.S. National Science Foundation through grant NSF-PHY-0244786 at the
University of Tennessee, and the U.S. Department of Energy under contract
DE-AC05-00OR22725 at Oak Ridge National Laboratory.

\end{document}